\documentclass[5p]{elsarticle}

\usepackage[mathscr]{euscript}

\usepackage{graphicx}
\usepackage[font+=footnotesize]{subcaption}

\usepackage{amssymb}
\usepackage{amsthm}
\usepackage{amsmath}
\usepackage{bm}

\usepackage{enumitem}

\usepackage{multirow}

\usepackage{listings}
\usepackage{xcolor}
\usepackage{bera}
\usepackage{hyperref}

\colorlet{punct}{red!60!black}
\definecolor{background}{HTML}{EEEEEE}
\definecolor{delim}{RGB}{20,105,176}
\colorlet{numb}{magenta!60!black}

\mathchardef\hyph="2

\lstdefinelanguage{javascript}{
  keywords={typeof, new, true, false, catch, function, return, null, catch, switch, var, if, in, while, do, else, case, break},
  ndkeywords={class, export, boolean, throw, implements, import, this},
  basicstyle=\scriptsize\ttfamily,
  numberstyle=\scriptsize,
  ndkeywordstyle=\color{darkgray}\bfseries,
  identifierstyle=\color{black},
  sensitive=false,
  numbersep=8pt,
  comment=[l]{//},
  morecomment=[s]{/*}{*/},
  frame=lines,
  backgroundcolor=\color{background},
  morestring=[b]',
  morestring=[b]"
}
 
\lstdefinelanguage{hyperflow}{
    keywords=[7]{functions, processes, signals, ins, outs, schemas},
    comment=[l]{//},
    basicstyle=\scriptsize\ttfamily,
    numbers=left,
    numberstyle=\scriptsize,
    stepnumber=1,
    numbersep=8pt,
    showstringspaces=false,
    breaklines=true,
    emphstyle=\ttb,
    frame=lines,
    backgroundcolor=\color{background},
    literate=
     *{0}{{{\color{numb}0}}}{1}
      {1}{{{\color{numb}1}}}{1}
      {2}{{{\color{numb}2}}}{1}
      {3}{{{\color{numb}3}}}{1}
      {4}{{{\color{numb}4}}}{1}
      {5}{{{\color{numb}5}}}{1}
      {6}{{{\color{numb}6}}}{1}
      {7}{{{\color{numb}7}}}{1}
      {8}{{{\color{numb}8}}}{1}
      {9}{{{\color{numb}9}}}{1}
      {:}{{{\color{punct}{:}}}}{1}
      {,}{{{\color{punct}{,}}}}{1}
      {\{}{{{\color{delim}{\{}}}}{1}
      {\}}{{{\color{delim}{\}}}}}{1}
      {[}{{{\color{delim}{[}}}}{1}
      {]}{{{\color{delim}{]}}}}{1},
}

\usepackage{newverbs}

\makeatletter
\newcommand\mv[1][white]{%
    \@testopt{\@mv{#1}}{-#1}
}
\def\@mv#1[#2]{%
    \Collectverb{\@@mv{#1}{#2}}%
}

\definecolor{light-gray}{gray}{0.93}

\def\@@mv#1#2#3{%
    \colorbox{light-gray}{\scriptsize \tt #3}
}

\journal{arXiv}

\begin{document}

\begin{frontmatter}





\title{Machine Learning in the Internet of Things for Industry 4.0}


\author{Tomasz Szydlo}
\author{Joanna Sendorek}
\author{Robert Brzoza-Woch}
\author{Mateusz Windak}

\address{AGH University of Science and Technology,\\
Department of Computer Science, Krakow, Poland}

\begin{abstract}
Number of IoT devices is constantly increasing which results in greater complexity of computations and high data velocity. One of the approach to process sensor data is dataflow programming. It enables the development of reactive software with short processing and rapid response times, especially when moved to the edge of the network. This is especially important in systems that utilize online machine learning algorithms to analyze ongoing processes such as those observed in Industry 4.0. In this paper, we show that organization of such systems depends on the entire processing stack, from the hardware layer all the way to the software layer, as well as on the required response times of the IoT system. We propose a flow processing stack for such systems along with the organizational machine learning architectural patterns that enable the possibility to spread the learning and inferencing on the edge and the cloud. In the paper, we analyse what latency is introduced by communication technologies used in the IoT for cloud connectivity and how they influence the response times of the system. Finally, we are providing recommendations which machine learning patterns should be used in the IoT systems depending on the application type.
\end{abstract}


\end{frontmatter}


\section{Introduction}
Due to the advances in electronics and computer systems, we are observing an increasingly large number of embedded devices connected to the Internet. Handling hundreds of thousands of interactions between such devices is a very challenging task and appropriate organization of data processing is necessary. There are several aspects that have to be dealt with in this context, including how, where and what computations should be performed.


IoT systems often assume the interaction of devices with cloud services which makes the Service Oriented Architecture (SOA) a suitable approach to designing processing logic. In that concept  
devices expose their features like services, which are then consumed by other services deployed in the cloud \cite{rajesh2010integration, brzoza2016fpga}. This concept has since evolved into the so-called Web Of Things \cite{Guinard:2016:BWT:3055920}. Nevertheless, the large volume of streaming data generated by IoT devices can make that approach inefficient. The solution to this problem is to make processing data-driven and responsive. Flow-based programming \cite{morrison2010flow} is a paradigm that defines applications in terms of flows of processes which exchange data by passing messages across predefined connections. The execution time of operations is dependent on their order in the processing flow. This concept can be adapted for IoT, where things are sources of data streams to be processed. There are several runtime environments that implement this concept, such as CppFBP \cite{morrison2010flow}, NoFlo \cite{bergius2015noflo}, WotKit \cite{blackstock2012iot} and NodeRED. 

Efficient flow processing of streaming data in IoT is especially important in the industry, currently undergoing revolutionary changes colloquially referred to as Industry 4.0 \cite{Hermann:2016}. The name refers to the notion of a fourth industrial revolution, where the focus shifts to delivering innovative services and products. Previous stages of the industrial revolution introduced mechanization through the use of (1) steam engines, (2) electricity (facilitating mass production), and (3) IT solutions. The fourth step is to use digital product models developed according to customer requirements and manufactured by Smart Factories \cite{Brettel:2014}. Such factories will be capable of self-planning and self-adaptation using devices compatible with the concept of the Internet of Things and Cyber-Physical Systems \cite{4519604}. 

For optimal production of goods and their subsequent delivery, Industry 4.0 proposes the use of machine learning mechanisms to study customer requirements and ensure optimal use of industrial equipment and to enforce predictive maintenance. Self-planning and self-adaptation processes require a collection of knowledge from the factories, development of models and applying them to obtain predictions. In such usage, short response times are necessary. For example, in factory automation or motion control response times lower than 10ms are mandatory, while process automation can be satisfied with 100ms response times. Processing flows which contain machine learning algorithms can be executed in the cloud, offering high processing capacity, a global view of the IoT system and easy management; however, this also introduces additional time delay as data has to be streamed from sensors to the cloud. Deploying such flows only in the fog, on the network edge, limits knowledge sharing because sensor data is not exchanged between remote premises but improves system response times. Lack of information exchange between branches may result in the situation that the system will not be able to recognize events that have already occurred elsewhere.

In the paper, we present the generalized concept of a flow-based processing stack and how the underlying infrastructure may influence dataflow execution. We propose three architectural patterns for machine learning algorithms that can be applied to data flows prior to their deployment and execution on real hardware. As a result, the computation is divided into a part located in the cloud, allowing for knowledge dissemination, and an edge part which provides high responsiveness. The proposed patterns provide a tradeoff between the accuracy of machine learning models and prediction times. In the evaluation of the concept, we are focusing on online supervised learning methods.
We argue that systems based on flow-based programming for IoT should take into account a holistic view of the system \cite{BALIS2018128} which perceive the system as a whole - in particular, the interplay of conflicting objectives and configuration options across all subsystems i.e. underlying hardware, communication topology, operating systems and execution environments. 
The scientific contribution of the paper includes (i) a generalized concept of a flow-based processing stack for IoT, (ii) machine learning architectural patterns for IoT, (iii) evaluation of the communication technologies used in the IoT systems for cloud connectivity, and (iv) recommendations which pattern and communication technology is appropriate for specific types of IoT systems. 

The structure of the paper is as follows. Section 2 presents the state of the art of machine learning algorithms for embedded devices. In section 3 dataflow execution is discussed. Section 4 discusses ML for IoT and proposes delta patterns. Section 5 contains an evaluation of the proposed concepts while section 6 sums up the paper.

\section{State of the art}
In large-scale computing, TensorFlow \cite{abadi2016tensorflow} is among well recognised neural network-based machine learning systems for distributed computing platforms. Due to multi-layer design, it runs on multiple hardware platforms. Its computing tasks are dispatched to many kernels running on many CPU and GPU cores. The lite version of the TensorFlow is a set of tools to run neural network models on mobile and embedded devices but the machine learning model has to be optimized to run efficiently on resource constrained devices. The first group of optimization techniques is related to the neural network accelerators. Lane et al. \cite{lane2016accelerated} present a software accelerator that enhances deep learning execution on heterogeneous hardware, including mobile devices. An original approach to machine learning in resource-constrained embedded systems is presented by Bang et al. \cite{bang201714} where authors describe a hardware accelerator for performing deep-learning tasks while requiring very low power to operate. The similar chips are designed by Intel and Google. The second group of solutions is related to the methods of neural network parameters quantization in order to diminish the amount of computation data storage and transfer time~\cite{BitFusion}. There are also works related to other machine learning algorithms. Shamili et al.~\cite{shamili2010malware} propose the utilization of Support Vector Machine (SVM) running on networked mobile devices to detect malware. Implementation of algorithms related to machine learning domain on extremely resource-constrained devices has also been described, e.g. in \cite{gupta2017protonn, kumar2017resource}. 

The increased interest in systems operating on the edge of the network has resulted in the research on machine learning solutions that combines processing on the network edge and in the cloud. The work of Liu et al. \cite{liu2017new} describes an approach for image recognition in which the process is split into two layers: local edge layer constructed with mobile devices and remote server (cloud) layer. In the edge, i.e. on a mobile device, an acquired image is preprocessed and segmentation is performed. Then the image is classified on a remote server running pretrained convolutional neural network (CNN). A more general survey on employing networked mobile devices for edge computing is presented in \cite{tran2017mobile}. There is also an ongoing research aimed at using machine learning methods to optimize energy consumed by the devices. Peralta et al. \cite{IndustryFog4} describe the method for sending sensor data from devices to the cloud that filters values which could be predicted by the machine learning methods.   



In Industry 4.0, due to internal processes such as machine ageing, an important aspect is the need for dynamic adaptation to new patterns in the data or when the data is generated as a function of time. This means that the aforementioned methods of moving learned machine learning models to edge devices are insufficient because over time they become obsolete. Models must be constantly updated to match the current nature of the process. A comprehensive survey on incremental online learning is presented in \cite{LOSING2017}. Yin et al. \cite{yin2016improved} present an application of incremental learning for object recognition. In contrast, \cite{provodin2016fast} describes an application of learned navigation system to control an autonomous robot. These publications, however, do not concern resource-constrained or embedded systems that operate at the network edge or in a fog environment. In this paper, we propose an organizational pattern for such machine learning algorithms on the edge and in the cloud and then discuss its usage taking into account the hardware capabilities of IoT systems and the available communication technologies.

\section{Dataflow execution}
\begin{figure}[!htbp]
  \centering
  \includegraphics[width=0.9\columnwidth]{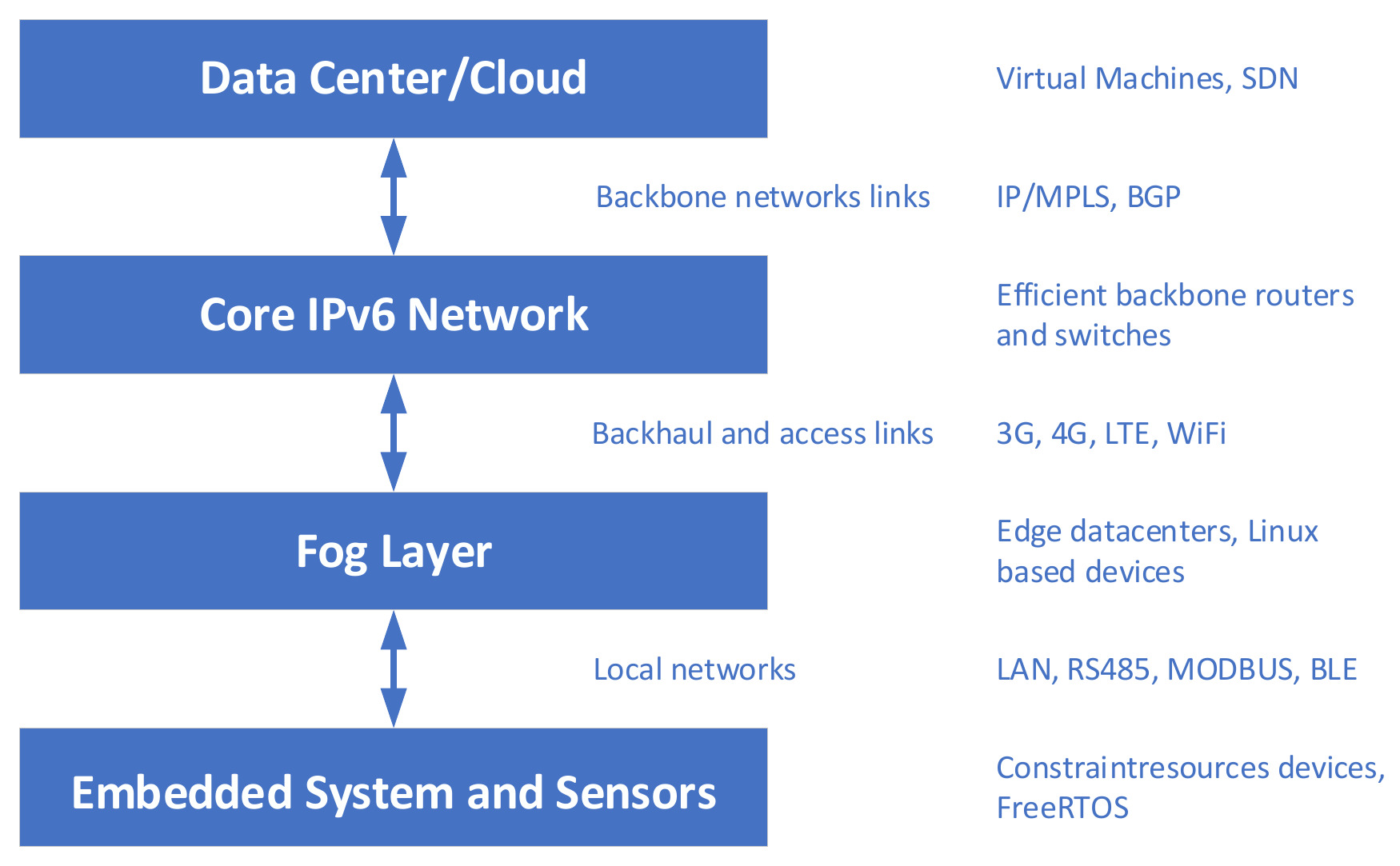}
  \caption{Fog computing architecture}
  \label{fig:hardware}
\end{figure}

The kind of systems discussed in the paper follows the concept of fog computing which was first described by Cisco \cite{bonomi2012fog, yi2015survey}. Fig.~\ref{fig:hardware} depicts the architecture of such systems. One of the most important factors for processing data in the IoT, especially for industrial IoT, is the fast response time. Table~\ref{figt:resp} contains the required latencies for critical IoT applications \cite{schulz17}. The processing flow can be executed in the cloud, providing virtually unlimited computational resources (e.g. Amazon AWS, Microsoft Azure or Google Cloud) and network bandwidth (e.g. fibre links in Core Internet), but the data transmission latency (e.g. via GSM) might be unacceptable for timely reception of results. On the other hand, executing data flows in the fog layer provides responsiveness but lacks a global view of data. This means that, for example, in a factory, the knowledge acquired during the processing of sensor data in one of its branches is not transferred to another branch. Because of this, the system may not detect potentially dangerous situations that have already occurred.

\begin{table}[!htbp]
\centering
\caption{Required latencies for critical IoT applications}
\begin{tabular}{c}
  \includegraphics[width=0.60\columnwidth]{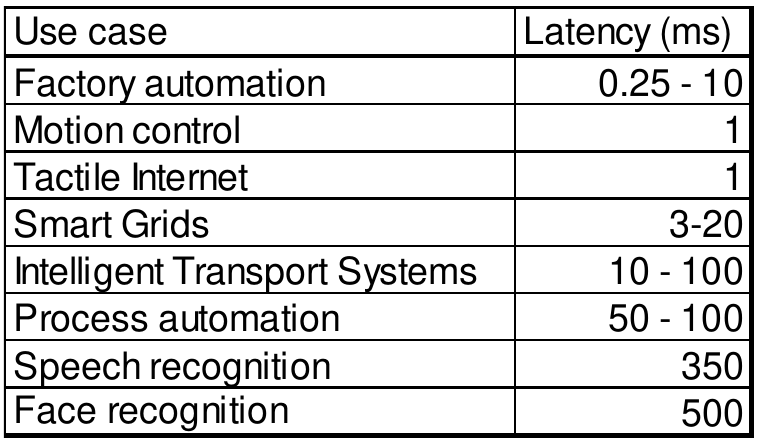}
\end{tabular}
\label{figt:resp}
\end{table}


\begin{figure*}[!htbp]
  \centering
  \includegraphics[width=0.7\textwidth]{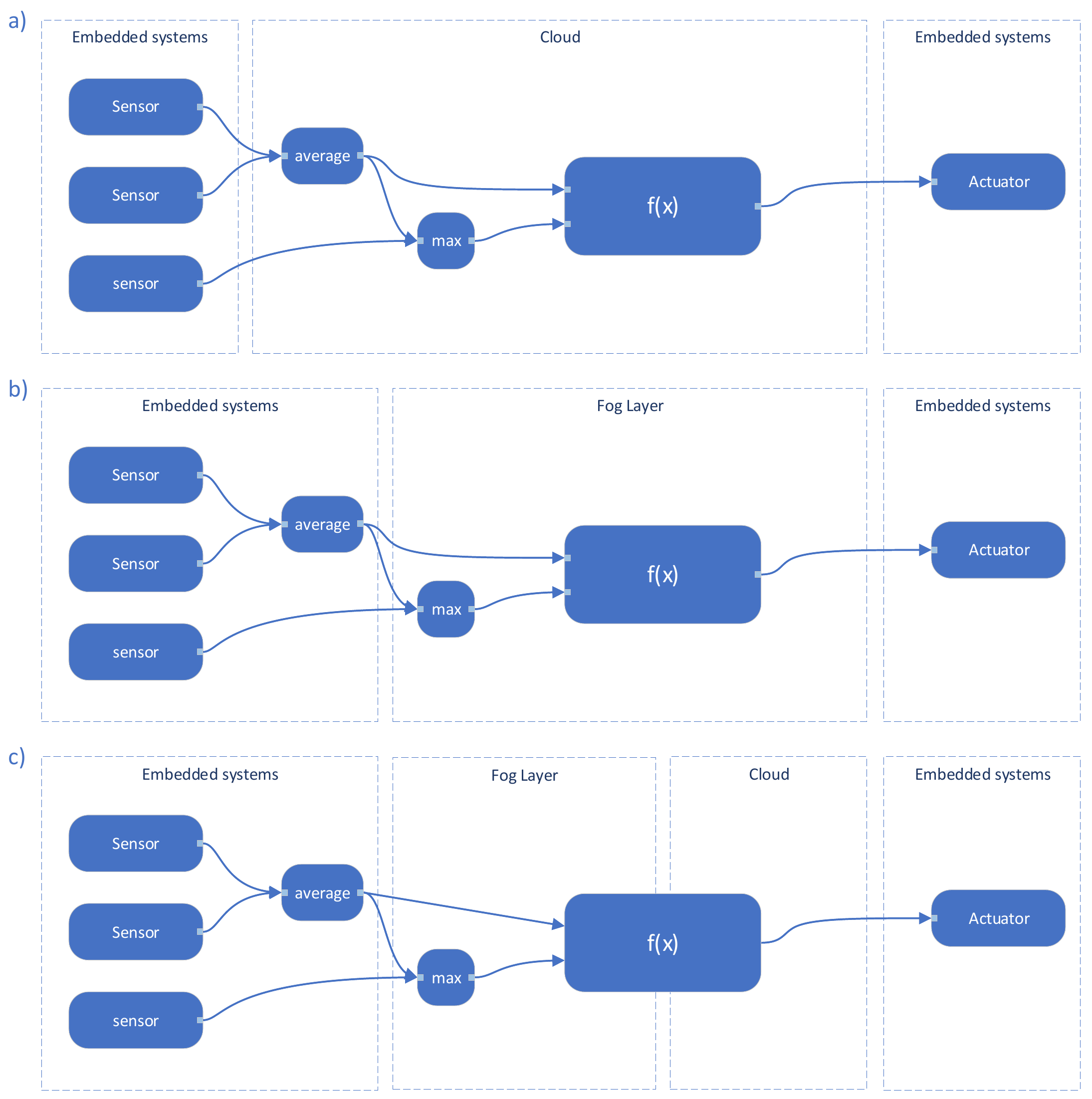}
  \caption{Flow execution examples}
  \label{fig:execution}
\end{figure*}

Fig.~\ref{fig:execution} depicts examples of flow execution where some elements are executed on different devices. Fig.~\ref{fig:execution}a depicts a situation where data streams from sensors are sent to the cloud for processing. This provides the possibility to collect data from thousands of sensors and process it at a single site, providing a high quality of results. On the other hand, such processing introduces huge latencies. In Fig.~\ref{fig:execution}b processing is performed locally on the edge, which ensures short processing times, but the information is not shared between edges (e.g. several factories located on different premises). This solution can be regarded as execution in a private cloud. Finally, Fig.~\ref{fig:execution}c shows a situation where the processing logic is spread between the fog and the cloud. Sensor data is preprocessed in the fog and only aggregated data is sent to the cloud. This reduces processing time and allows for knowledge sharing between geographically distributed devices deployed in the fog. Machine learning algorithms can be used to update models in the cloud, and then move them to the fog on a regular basis. This represents a middle ground between the previously discussed cases.

\begin{figure}[!htbp]
  \centering
  \includegraphics[width=0.8\columnwidth]{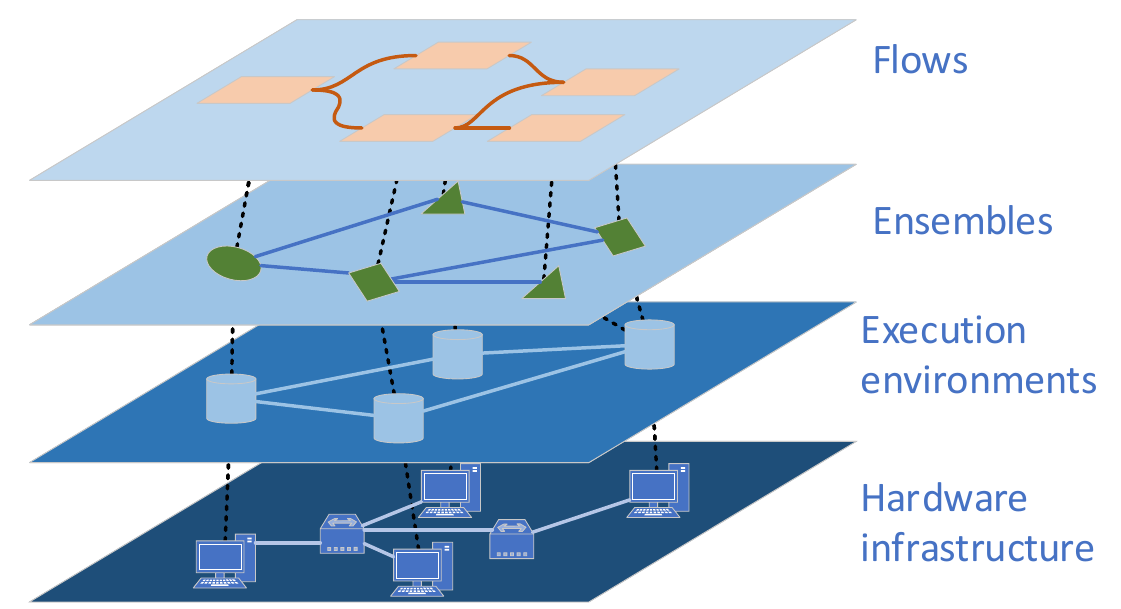}
  \caption{Flow Processing Stack for IoT}
  \label{fig:stack}
\end{figure}

The usability of the aforementioned deployment models depends, among others, on the type of devices used in the fog layer, the capability for performing learning processes, and communication technologies. Thus, flow-based processing for IoT may be analyzed in the context of the Flow Processing Stack (FPS). FPS is a generic model suited for IoT solutions. It is conceptually presented in Fig.~\ref{fig:stack}. The stack consists of four layers: 

\begin{itemize}
\item \textbf{Flows} -- A data flow can be described as a directed graph, where vertices represent computational processes while edges define the flow of messages between them. It represents a computation from the business logic perspective without dealing with technological aspects related to data serialization and transmission. 
\item \textbf{Ensembles} -- Ensembles are the executable elements that can be deployed and executed in the execution environments. Depending on their type, they can assume the form of JSON files for \textit{NodeRED} software, generated source code that can be executed in virtual machines, lightweight containers described by dockerfiles or firmware for an embedded processor. 
\item \textbf{Execution environments} -- An execution environment represents a particular execution platform, such as an operating system or an application container that manages the life-cycle of the application. Execution environments are typically part of other computing hardware or systems. Execution environment should provide operating system-level services, necessary software libraries, required memory and processing power as needed. In our case, the execution environment is a flow-based processing engine such as \textit{NodeRED}, \textit{uFlow} \cite{szydlo:2017} or others. 
\item \textbf{Hardware infrastructure} -- represents the hardware infrastructure upon which the execution environments and ensembles are executed. It covers not only the devices and their capabilities but also the network topology and communication technologies.
\end{itemize}

High-level processing flows have to be transformed into a set of ensembles that can be deployed in the execution environments. For example, data flow can be divided into several sub-flows, where some will be executed in the \textit{NodeRED} execution environments installed on various Linux devices and others on \textit{uFlow} executed on embedded devices. 

In the paper, we focus on machine learning algorithms and how to organize the learning and predicting processes. In the next section, the concept of machine learning architectural patterns will be discussed. Later on, we analyze latencies introduced by common technologies in the network core and the backhaul that may have an impact on latency, thus determining the applicability of the system.

\section{Machine Learning for IoT}
The classic approach to machine learning for IoT assumes learning using cloud computing (referred to as Big Data ML). Data from sensors is transmitted to the cloud for analysis over the Internet. Access to historical data stored in a central repository has a number of advantages as it enables the development of appropriate machine learning methods to match the needs of specific applications. Unfortunately, centralized data processing also suffers from a number of drawbacks:
\begin{itemize}
\item the ever-growing number of devices in the Internet of Things generates an increasing amount of data that must be transmitted and stored in the cloud;
\item lack of autonomy in local subsystems (due to the requirement to transfer data to the cloud for processing) means that such subsystems cannot perform their function in the absence of network connectivity;
\item online systems are usually slow to respond to new events as they must first upload data to the cloud (which introduces delays);
\item there are consumer concerns related to the protection of private data when transferring data to the cloud or asking for personalized results.
\end{itemize}

Moving some of the machine learning computations closer to sensors, i.e. to the network edge (referred as Local ML) should be transparent to machine learning algorithms providing results comparable (in terms of quality) with the Big Data ML approach while retaining the scalability, personalization, confidentiality, responsiveness and autonomy characteristic of the Local ML approach. 

In contrast to the stationary problems that can be solved offline using machine learning algorithms, in the IoT domain most of the problems are related to data streams. These are generated by a number of sensors deployed in real-world devices. One of the important problems related to such data is concept drift \cite{drift:1,drift:2}. It refers to a phenomenon where the target concept changes over time. For example, the behaviour of customers at a shop may evolve, while machines deployed in factories may change their vibration characteristics due to the wear of their components. These problems can be handled by online learning algorithms, where 
the model is continuously updated with new data. We distinguish two main groups of such algorithms -- those that use new data to adjust internal models and those that retrain themselves every time new data is received. The former group does not need to keep all training data in memory, so it is better suited to streaming data, while the latter stores all training data. Losing et al. analyzed various incremental online learning algorithms \cite{LOSING2017} and discussed their applicability to data characterized by concept drift. 

\begin{figure}[!htbp]
  \centering
  \includegraphics[width=0.35\columnwidth]{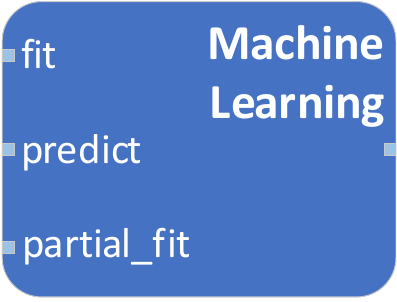}
  \caption{Machine learning processing block}
  \label{fig:ml_block}
\end{figure}

Fig.~\ref{fig:ml_block} depicts the machine learning block that can be used in the dataflow. The \textit{fit} input trains the model based on the provided data, \textit{partial\_fit} corrects internal models using new data and, finally, \textit{predict} uses the internal model to estimate an output value. In the following subsection, we discuss the internal structure of that block, mindful of the \textit{delta architecture} concept.

\subsection{Delta architecture for machine learning}
Delta architecture represents a framework for machine learning algorithms where calculations are spread on the network edge and the cloud. Two distinct architectural layers are defined:
\begin{itemize}
 \item Collective Intelligence Layer/Central Layer based in the cloud and used to analyze and process data representing the so-called collective intelligence. This layer enables the discovery of general relationships and trends present in centrally stored data. It can also use knowledge from services available on the Internet.
 \item Edge Layer comprising devices with limited resources located on the network edge and usually operating as the Internet gateway for IoT devices. This layer is capable of online sensor data processing. Due to the limited capabilities of its component devices, the volume of data that can be processed and stored locally is limited.
\end{itemize}

\begin{figure*}[!htbp]
  \centering
  \includegraphics[width=0.95\textwidth]{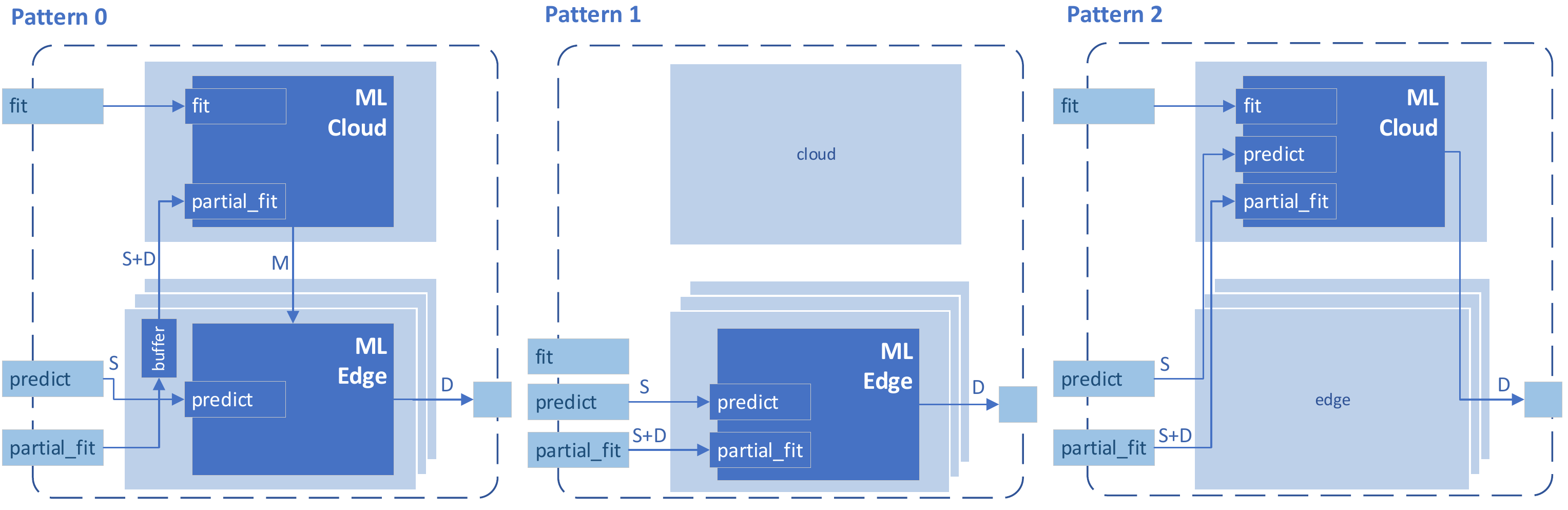}
  \caption{Delta Patterns}
  \label{fig:patterns}
\end{figure*}

Fig.~\ref{fig:patterns} depicts the delta patterns for machine learning algorithms. We focus on situations where predictions using estimators should be performed on the edge. In our patterns we distinguish four types of messages:
\begin{itemize}
 \item{S} - sensor data;
 \item{S+D} - sensor data and the predicted value;
 \item{M} - serialized machine learning model;
 \item{D} - prediction/decision provided by the model.
\end{itemize}

The following subsections present three architectural patterns of the machine learning processing block.

\subsection{Pattern 0}
In pattern P0, data from sensors are aggregated and transmitted to the cloud where BigData ML mechanisms generate global models of behaviour based on the received data. The central layer processes data collected from a number of devices and sites. It can adapt to various conditions and creates general models which are then distributed to the edge layer. Predictions performed in the edge layer, based on the sensor data, can be made locally ensuring short response times. Aggregated sensor data can be sent to the cloud at an appropriate time optimizing resource consumption \cite{FedCSIS2014252}. In this pattern, learning is only performed in the cloud layer, while predictions which apply the model occur on the edge.

\subsection{Pattern 1}
In the edge learning pattern, sensor data is processed locally by machine learning modules at the edge. On this basis, a local machine learning model appropriate for a limited area is trained. Knowledge is not exchanged across edge locations. In this pattern, machine learning models used on the edge can be set in advance based on offline learning using historical data.

\subsection{Pattern 2}
The cloud learning pattern uses the mechanisms of machine learning in the cloud layer. Data from sensors is transmitted to the cloud where BigData ML mechanisms generate global models of behaviour based on the received data. Predictions based on the acquired knowledge are performed in the cloud. The results are accurate but these patterns introduce the highest processing latency.

\section{Evaluation of delta patterns for machine learning}
In this section, we present an experimental evaluation of the proposed mechanisms. Our goal was twofold: (i) measure how the delta patterns influence the accuracy of the common ML models and (ii) determine what prediction latency can be achieved using different back-haul technologies. Comparing both the achieved prediction accuracy and latency using different patterns allow suggesting pattern suitable for the given case as presented in section \ref{summary}.

For the purposes of the case study, we prepared synthetic data that represent typical situations observed in the industry. We assumed a single cloud processing infrastructure and five independent edge environments, e.g. five premises with newly installed product lines or pumps. These devices are equipped with several sensors that generate streams of data representing the internal parameters of each device. All devices face the same problem, i.e. the ageing of their components. Machine learning algorithms are used to predict the state of the machine -- from problem-free operation through various warning states all the way to failures. 


The following subsections analyze the hardware infrastructure of devices that might be deployed on the edge, machine learning algorithms and, finally, backhaul communication technologies.

\subsection{Hardware infrastructure analysis}
Our case study involves four hardware infrastructures, as presented in Table~\ref{figt:casestudy}. The table shows how processing flows can be mapped to ensembles for a given hardware infrastructure which is composed of the three elements, i.e. sensors, gateways and the cloud (referring to the aforementioned fog computing concept).

\begin{table*}[!htbp]
\caption{Case study description}
\centering
\begin{tabular}{c}
  \includegraphics[width=0.95\textwidth]{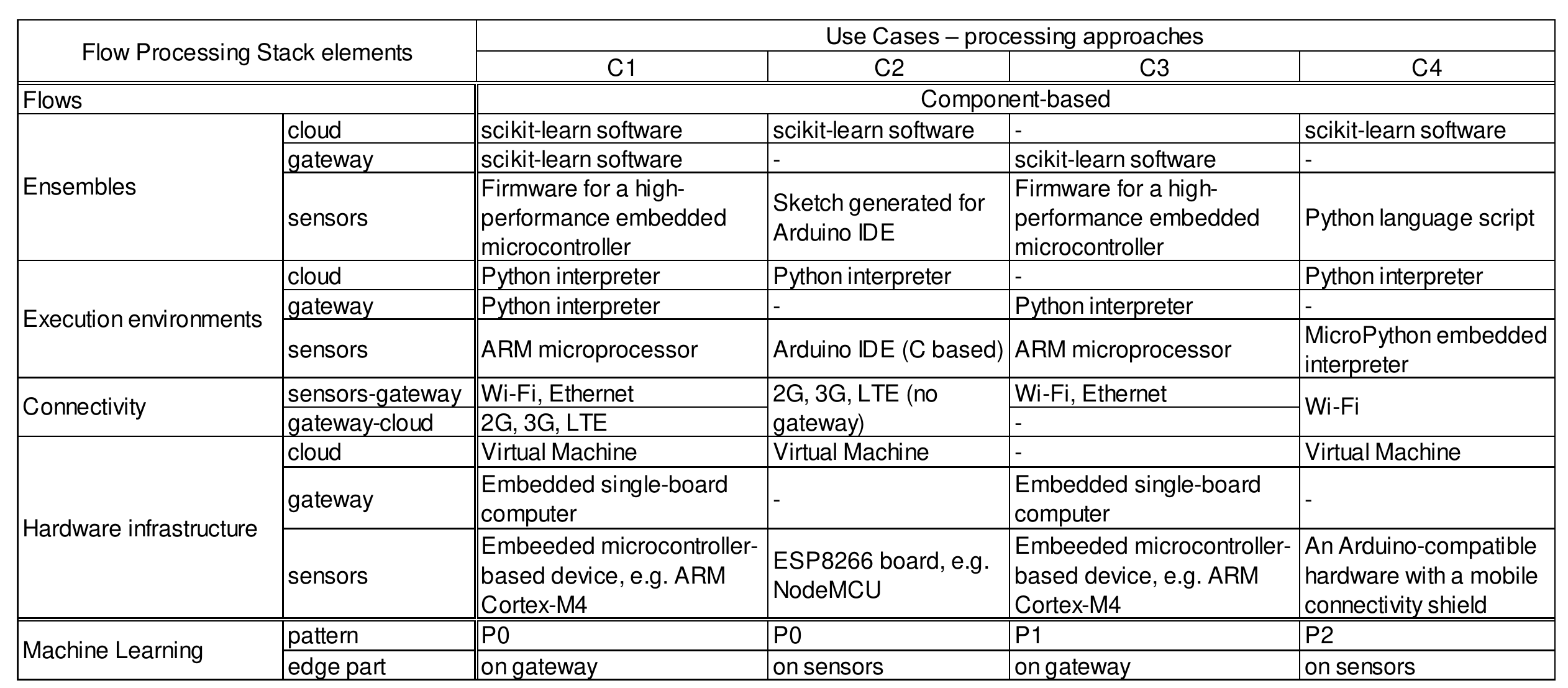}
\end{tabular}
\label{figt:casestudy}
\end{table*}


C1 and C2 are the representation of the sample processing flow from Fig.~\ref{fig:execution}c. In both cases the proposed delta pattern is P0. In C1, the sensors send data to gateways which transmit it to the cloud. The pattern then determines whether to send the updated model to the edge. The machine learning algorithms on the edge are deployed on gateways. We have analyzed various ML libraries, such as Weka, Moa and Python Scikit-learn, and finally decided to use the latter in our experiments. Our decision was dictated by the fact that Python is a commonly used programming language for IoT gateways. 

In C2, the sensors equipped with GSM connectivity are able to send data directly to the cloud. Since sensors have limited resources, they cannot directly run a Python library used previously to train ML models. In \cite{codegen} we have proposed the solution that converts the ML model to source code and then compiles it into device firmware. Table~\ref{tab:conversion} shows how the \textit{scikit-learn} decision tree model can be transformed to C source code for an embedded microprocessor using the library FogML\footnote{https://github.com/tszydlo/FogML accessed 04.10.2019}. It is interesting to note that the Python model serialized using the Python serialization library called \textit{pickle} is 2686 bytes long, while the binary firmware of the C code of the \emph{predict} function compiled using \textit{arm-none-eabi-gcc} is only 523 bytes long.

\begin{table}[]
\tiny
\centering
\caption{Machine learning model transformation}
\label{tab:conversion}
\begin{tabular}{|p{8cm}|}
\hline
\textbf{Python scikit-learn decision tree model}
\\ \hline
\begin{lstlisting}[]
node=0 test node: go to node 1 if X[:, 0] <= 3.881 else to node 12.
    node=1 test node: go to node 2 if X[:, 0] <= -0.185 else to node 7.
        node=2 test node: go to node 3 if X[:, 1] <= -1.662 else to node 4.
            node=3 leaf node.
            node=4 test node: go to node 5 if X[:, 1] <= 0.975 else to node 6.
                node=5 leaf node.
                node=6 leaf node.
        node=7 test node: go to node 8 if X[:, 1] <= 0.637 else to node 9.
            node=8 leaf node.
            node=9 test node: go to node 10 if X[:, 1] <= 4.341 else to node 11.
                node=10 leaf node.
                node=11 leaf node.
    node=12 leaf node.
\end{lstlisting}

\\
\hline
\textbf{Decision tree estimator for embedded processor}
\\ \hline
\begin{lstlisting}[]
int predict(float* x){
  if (x[0]< 3.88166737556) {                     //node = 0
        if (x[0] <= -0.185908049345) {           //node = 1
            if (x[1] <= -1.66271317005) {        //node = 2
                return 3;
            } else {
                if (x[1] <= 0.975374698639) {    //node = 4
                    return 5;
                } else {
                    return 6;
                }
            }
        }else {
            if (x[1] <= 0.637046217918) {        //node = 7
                return 8;
            } else {
                if (x[1] <= 4.3416762352) {      //node = 9
                    return 10;
                } else {
                    return 11;
                }
            }
        }
  } else {
      return 12;                                 //node = 12
  }
} 
\end{lstlisting}
\\
\hline
\end{tabular}
\end{table}

C3 refers to the flow depicted in Fig.~\ref{fig:execution}b. In this case, the machine learning model is trained only on the edge, on the gateway device. The recommended pattern is P1, where learning is performed on the data available in the particular edge environment. Of course, it is also possible to train ML models using BigData algorithms on historical data and then embed the final model in the device. 

C4 refers to the flow depicted in Fig.~\ref{fig:execution}a. Currently, this is a typical approach used in IoT solutions that rely on machine learning. The model is trained in the cloud on high-performance infrastructure and exposed as a service for devices located at the network edge. Nevertheless, the data from sensors has to be sent to the cloud and predictions have to be sent back to the sensors.

Integration of the ensembles deployed in various execution environments on different devices might be achieved using one of the IoT communication protocols \cite{surveyiot}. Previously we had been relying on the MQTT protocol for that purpose \cite{szydlo:2017}.

\subsection{Data sets with concept drift}
 In our experiments, we use stream datasets affected by concept drift which reflects various physical phenomena. We have two different datasets:
\begin{itemize}
\item \textbf{Circles} -- dataset consisting of points in two-dimensional space divided into seven categories of five hundred points each. The concept drift in this model is represented by constant-speed rotation by $\frac{\pi}{720 \times 3}$ per iteration. Each position change, which produces a new point in the stream, causes entire clusters to move.

\item \textbf{RandomTree} -- dataset generated with the use of the MOA data tool\footnote{https://moa.cms.waikato.ac.nz/ accessed 04.10.2019}. To emulate concept drift, the data stream comes from two RandomTree generators with different seeds, used in an alternating fashion. It is worth noting that this kind of concept drift is very slow and weak.
\end{itemize}

In the experiments, we analyze two scenarios. First, all of the symptoms and failures are observed at each premise. In the second scenario, at each premises one failure category is not observed, but may appear and should be detected -- e.g. a failure that has already appeared at other facilities. The purpose of the second scenario is to determine how knowledge can be shared between edges via a central processing module in the cloud. Thus, the stream data expressing physical phenomena is divided into five streams for each edge premises. We use two division methods when preparing test data:

\begin{itemize}
\item \textbf{equal} -- initial data stream is divided in such a way that each category is uniformly distributed through the edge environments,
\item \textbf{without-one} -- stream division causes one category to be missing in each edge environment.
\end{itemize}

The without-one method was used in order to simulate the situation when none of the edge devices has full knowledge of the data set characteristics. In this case, only the global model trained on the data coming from all of the devices has the ability to recognize all categories. 

\subsection{Evaluation of machine learning patterns}
We analyzed several machine learning algorithms and finally decided to use the Decision Tree Classifier, Support Vectors Machine with RBF kernel and Gaussian Naive Bayes. One of the deciding factors was the processing time of algorithms on embedded devices.

Due to the fact that the data stream is affected by concept drift, the trained model has to follow the changes. One of the possible solutions is to train models in a moving frame regime. In this case, the oldest data is dropped from the frame and new data is inserted as soon as it appears in the data stream. In our experiments, we decided to retrain ML models whenever the content of the moving frame changes. At each edge facility, the procedure applied when a new point arrives in the stream is as follows:
\begin{enumerate}
\item The point is classified with the currently assembled learning model. A score is calculated, which evaluates to 0 if the point has been misclassified and 1 otherwise. The average score for the most recent fifty samples is calculated to evaluate the accuracy of the model.
\item The point is added to the training set (moving frame) located in the cloud (pattern P0 or P2) or locally (pattern P1) only in the \textit{equal} division method scenario or when it is not selected to be abandoned in the \textit{without-one} scenario for the particular edge. 
\end{enumerate}

\begin{table*}[!htbp]
\caption{ML Patterns evaluation - P1 and P2}
\centering
\begin{tabular}{c}
\includegraphics[width=0.80\textwidth]{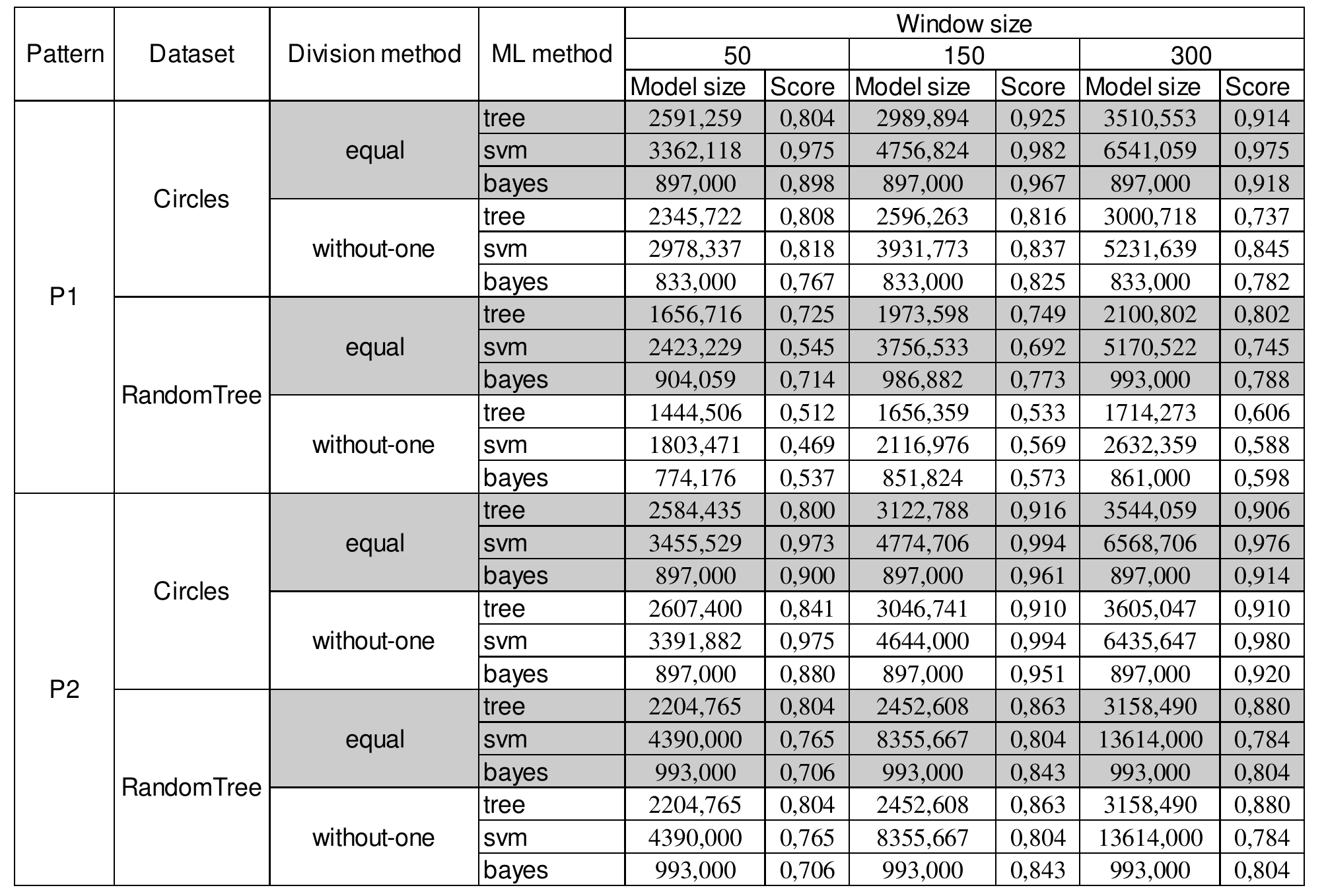}
\end{tabular}
\label{figt:patterns_p1_p2}
\end{table*}

\begin{figure*}[!htbp]
 \centering
 \includegraphics[width=1.0\textwidth]{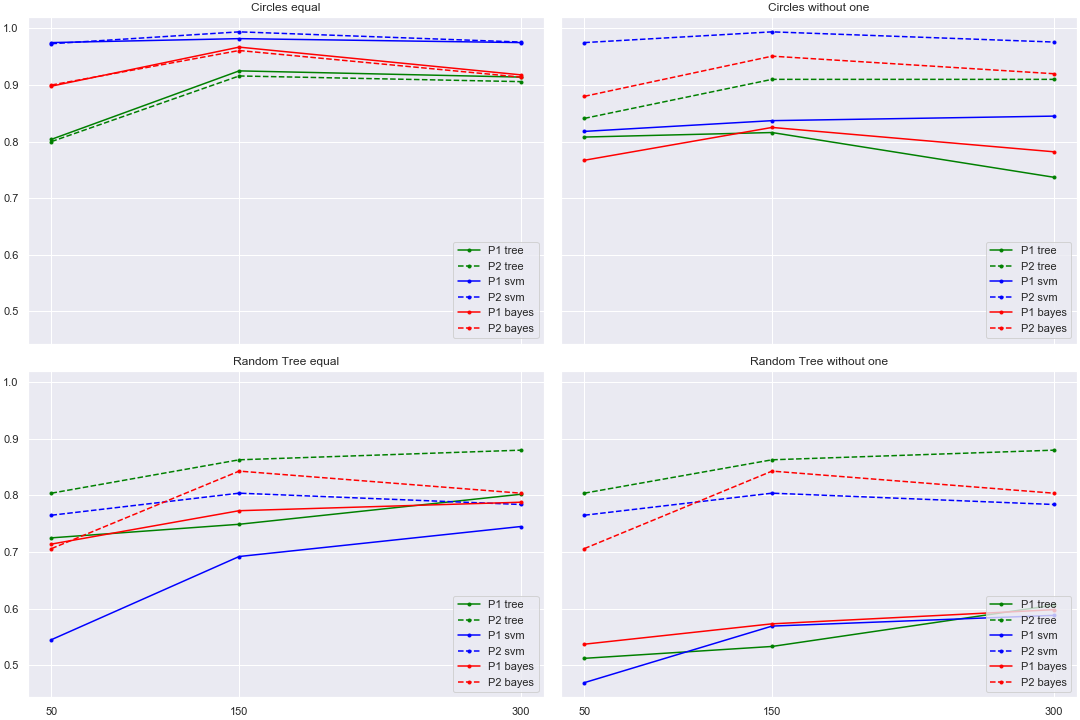}
 \caption{Summary of the evaluation of patterns P1 and P2 for RandomTree and Circles datasets.}
 \label{fig:charts}
\end{figure*}

Table \ref{figt:patterns_p1_p2} presents results achieved in experiments conducted for patterns P1 and P2. The size of the window on which the model is trained was 50, 150 and 300 respectively. For each case, the average model size and average score were calculated for the next 50 consecutive samples. Fig.~\ref{fig:charts} presents charts summarizing gathered data.

The following main conclusions can be drawn:

\begin{itemize}
\item For pattern P1 and both datasets, the \textit{equal} division case yields clearly better results than the \textit{without-one} case. This occurs due to the fact that storing one common training model in the cloud allows gathering more knowledge of data coming from many categories. 
\item For pattern P2 and both datasets, the results were similar regardless of the division method. This is due to the fact the machine learning is performed centrally in the cloud.
\item For data characterized by stronger concept drift (Circles dataset), moderate frame sizes yielded the best results. Lower scores for shorter frames are caused by the training set being too small, while longer frames cause data to retain greater concept drift and therefore reduced separability. 
\end{itemize}

Table \ref{figt:patterns_p0} presents results obtained for pattern P0 when the edge model changes every 150 iterations. The model score was calculated for the first 50 classifications after each model change, and similarly for further classifications. It can be noted that two datasets behave very differently depending on the frame offset after the model change. For the dataset burdened with higher concept drift (Circles dataset), results obtained during the first fifty iterations after the change are significantly better than in the following two frames. This is clearly caused by relying on more up-to-date training data, reducing the effects of concept drift. The greater the offset from the model change, the less up to date the model is. For the dataset with a lower concept drift (RandomTree), this phenomenon cannot be observed. It is also worth noting that model size remains constant regardless of the offset chosen. 

\begin{table*}[!htbp]
\caption{ML Patterns evaluation - P0}
\centering
\begin{tabular}{c}
\includegraphics[width=0.70\textwidth]{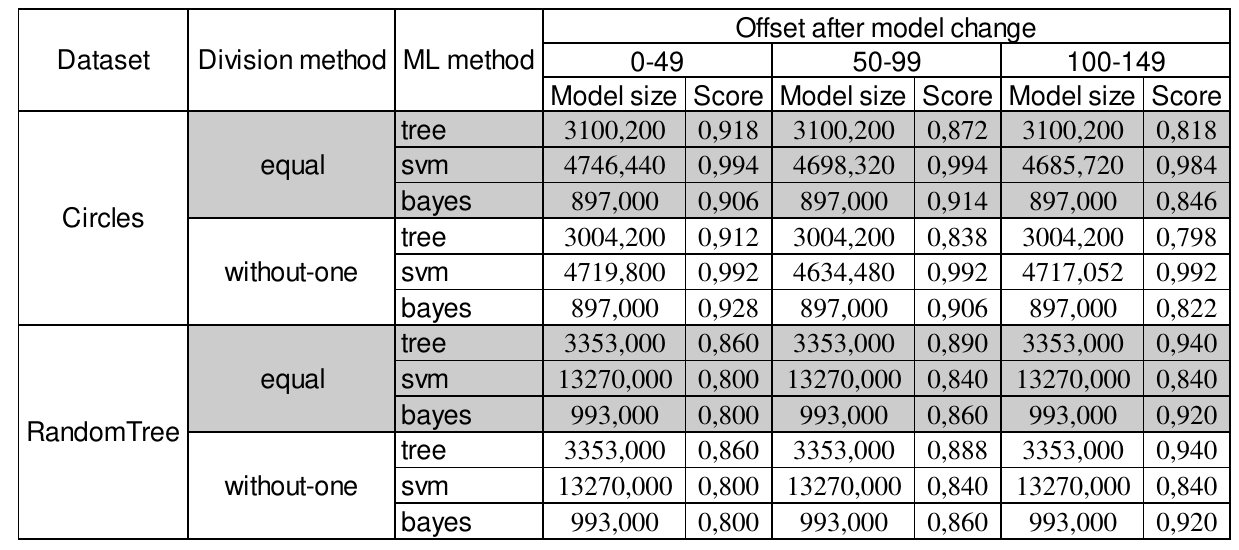}
\end{tabular}
\label{figt:patterns_p0}
\end{table*}

It is interesting to note that patterns P1 and P2 yield better results than P0 for the \textit{equal} division method of the data streams i.e. when knowledge transfer is not necessary. In the second case with the \textit{without-one} division method, P2 gives more accurate results than P0, and P0 gives more accurate results than P1. In P0 and P1 the decision made by the machine learning model is reached directly on the edge; thus response times are significantly lower than in P2 where data has to be sent to the cloud.

Our experiments show that it is possible to use machine learning algorithms in edge computing for non-stationary phenomena such as ageing of machines, and achieve short response times with moderate accuracy. The final results depend strongly on data distribution on the network edge. When processes are similar in each independent edge environment, local edge processing is sufficient. Nevertheless, when the edge environments need to exchange knowledge, cloud-based machine learning becomes mandatory. We have also shown that estimators trained using high-level libraries can be transformed into source code for embedded processors that can be flashed to their internal memory. By applying technologies such as OTA (over-the-air-programming) these processors can be reprogrammed at runtime. 

\subsection{Analysis of communication technologies}
Data transmission capabilities play an important role in the operation of multi-level edge-cloud architectures. We have evaluated several transmission media in terms of applicability to the use case of transferring ML models and control-measurement data. We chose media commonly utilized and accepted in the industry: IEEE 802.3 100BASE-TX Ethernet, 802.11 Wi-Fi, as well as 2G and 3G mobile communication.

The conducted tests evaluate the applicability of each medium for the C1-C4 use cases and the corresponding P0-P2 Delta Patterns described in this paper. The data transmitted and received during the tests is synthetic but representative of actual scenarios in which sensor data or ML models are sent over a selected medium.

\begin{figure}[!htbp]
  \centering
  \includegraphics[width=0.87\columnwidth]{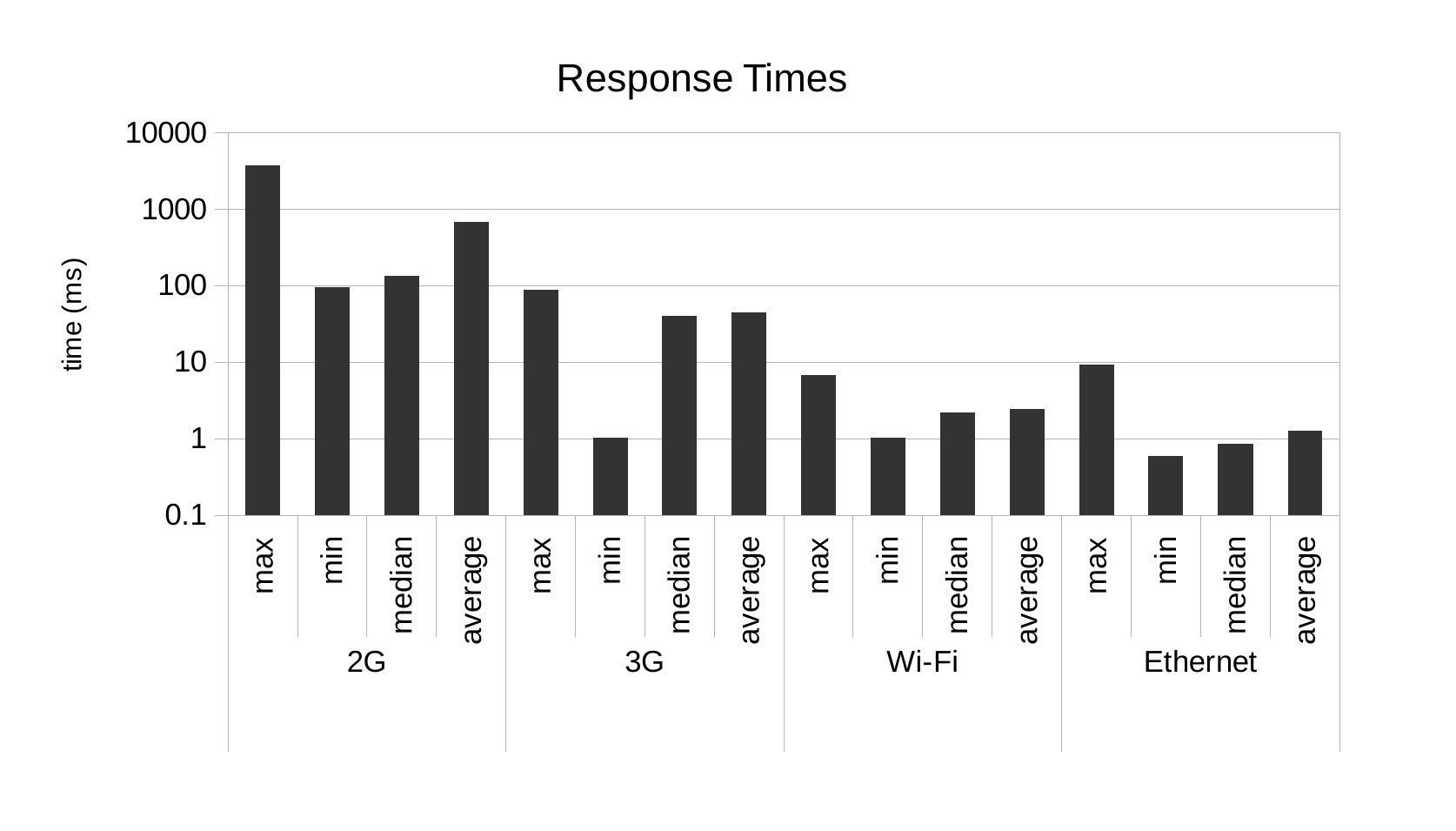}
  \caption{First hop response time for selected communication media.}
  \label{fig:resptime}
\end{figure}

As previously mentioned, adequate communication response times depend on the purpose of each system. Overall communication efficiency depends on several factors, most importantly network response time and transfer speed (transmission time of a given amount of data). The energy efficiency of the edge-layer nodes can be equally important, because, depending on the use case, these nodes may be powered with renewable sources. First, we evaluated and compared the response times for selected communication standards by analyzing the first hop response time using the traceroute tool. Results are shown in Fig.~\ref{fig:resptime}.

A rather obvious factor which significantly affects the response time of a connected system is the overall transmission delay. Further on in this paper, the entire transfer procedure will be referred to as a \emph{transaction} while the data transfer will be called \emph{data transmission}. For small chunks of data and responsiveness analysis, the average transmission speed given in e.g. bits/s is less important than the overall transaction time. This is why a second evaluation was also performed, focusing on the transmission times for four different chunks of data that correspond to typical use cases in lower layers of the proposed systems. The test data was transmitted between the communication device under test (DUT) and a remote server using a TCP-based protocol. Thanks to custom measurement hardware we were able to separate the connection set-up time (i.e. setting up a connection, disconnecting) and actual data transfer.

\begin{figure}[!htbp]
  \centering
  \includegraphics[width=0.99\columnwidth]{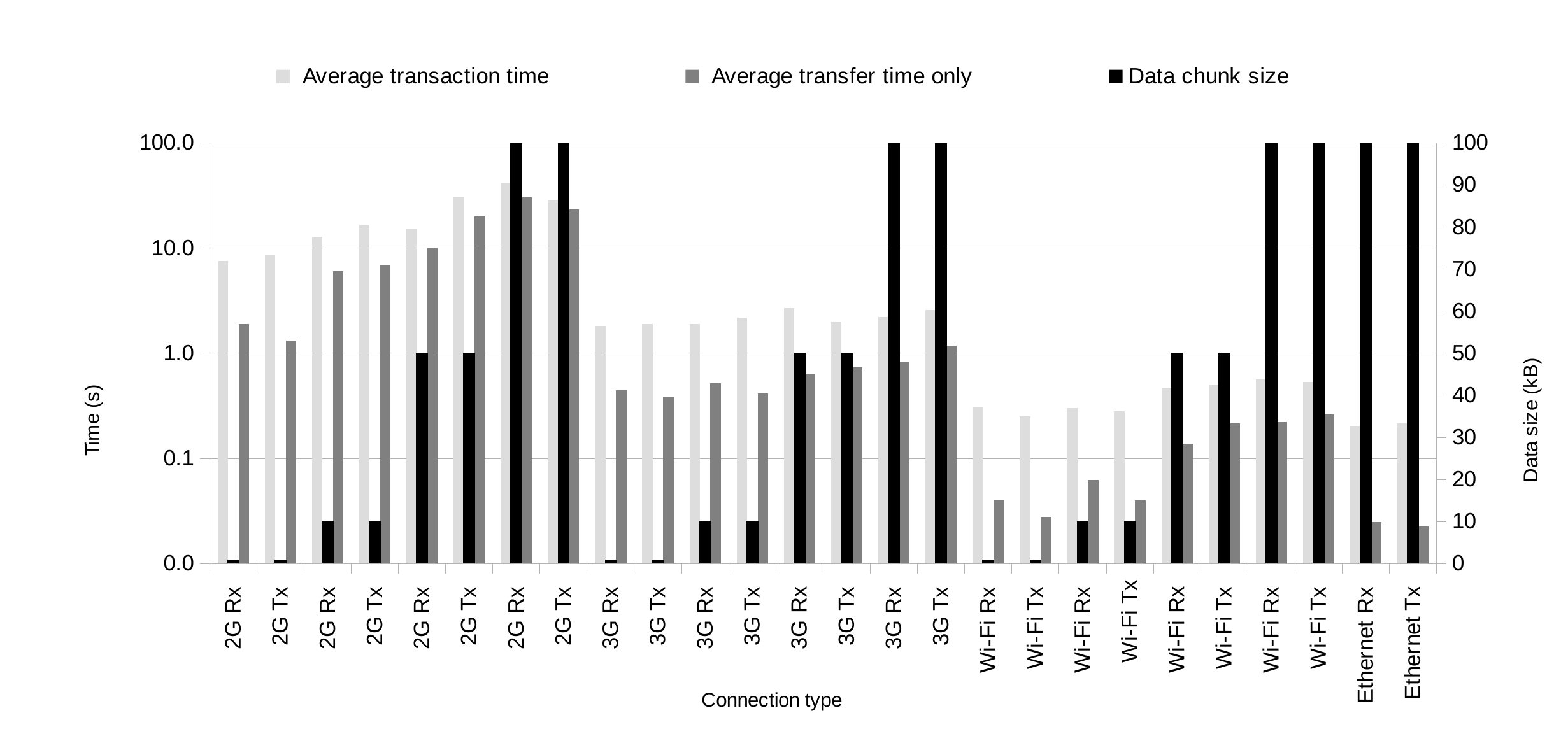}
  \caption{Transmission time for selected communication media.}
  \label{fig:transfertime}
\end{figure}

\begin{figure}[!htbp]
  \centering
  \includegraphics[width=0.99\columnwidth]{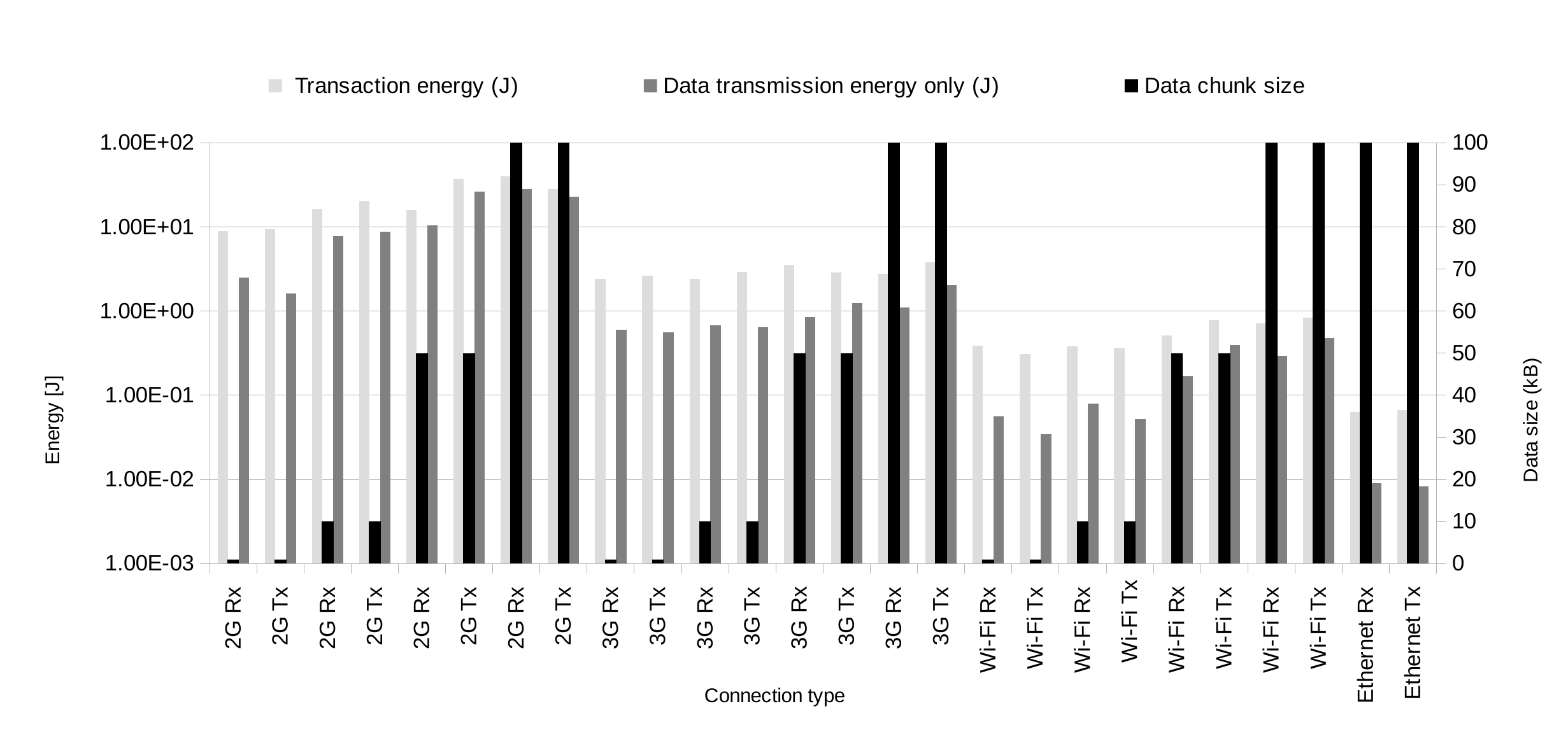}
  \caption{Energy requirements for selected communication media.}
  \label{fig:transferenergy}
\end{figure}

Results of time measurement are presented in Fig.~\ref{fig:transfertime}. The connection set-up time can be very significant, especially in slow networks, e.g. in 2G mobile communications. Ethernet network tests results are shown for the maximum chunk size only due to its very low transmission time (resulting in high speed) and insignificant increase in power consumption during transmission compared to an idle state. An additional but important factor concerns the energy efficiency of the communication standard. Fig.~\ref{fig:transferenergy} presents the comparison of energy requirements for the same tests as in Fig.~\ref{fig:transfertime}.

Many industrial connected devices such as telemetry appliances or automated environmental sensors utilize the well-established 2G mobile communication network. The standard can be used when the local system is deployed in a place which lacks a direct Internet connection or more advanced (3G, LTE) mobile network coverage. As can be noted in the presented evaluation, the responsiveness of the 2G network is relatively poor because the transaction time in every 2G transfer is longer than one second. This delay would make 2G unsuitable for modern industrial requirements (refer to Table~\ref{figt:resp}). As in this case computing power is much less expensive than data transmission, an obvious need arises for more advanced computing mechanisms which do not require any intensive communication with remote cloud servers. 

If a relatively slow connection is available, we propose the utilization of Delta Pattern P0 or P1 in which local processing plays a more important role than the available connection. Patterns 0 and 1 can also be applied to devices which work in low-rate networks, such as ZigBee. Additionally, it can be noticed that in P0 a trained model can be transferred to a local device in one of its background tasks, hence transmission time becomes a less important issue. If a direct Internet connection is available, the use of Ethernet or Wi-Fi network is strongly encouraged. This opens new possibilities regarding the availability of critical IoT applications (see Table~\ref{figt:resp}).

Another import aspect related to network communication is access to cloud resources using backbone links. We evaluated the latency of several Amazon Web Services (AWS) platform instances, obtaining results between 45ms for the closest AWS location up to 350ms for farther locations. Based on these results we may choose the best cloud provider location, offering the shortest latency. The overall latency for sending data from the fog to the cloud is the sum of latency introduced by the backhaul network technology and the time necessary to send data over the core Internet links.

\subsubsection{Evaluation summary} \label{summary}

To conclude, we assigned a recommended connection type for each of the previously mentioned application areas. A summary is presented in Table~\ref{figt:summary}. As can be noted, the use of the proposed patterns enables system designers to consider less demanding connectivity for applications that would require a faster and more expensive connection.

The main differentiating factor is the need for knowledge transfer in the cloud between edge environments. This is expressed in the table as \textit{edge/cloud learning} and \textit{edge learning only}. Another interesting observation is that for applications where response times greater than 100ms are acceptable, learning and predicting in the cloud is sufficient even for the edge learning case where there is no direct need to transfer knowledge. Of course, in the latter case, the edge environment is not autonomous and a persistent Internet connection is mandatory. 

\begin{table}[!htbp]
\caption{Summary of the recommended connection types for each of the described delta patterns (letters denote connection type: A: Ethernet, B: Wi-Fi, C: 3G or LTE, D: 2G)}
\centering
\begin{tabular}{c}
\includegraphics[width=0.95\columnwidth]{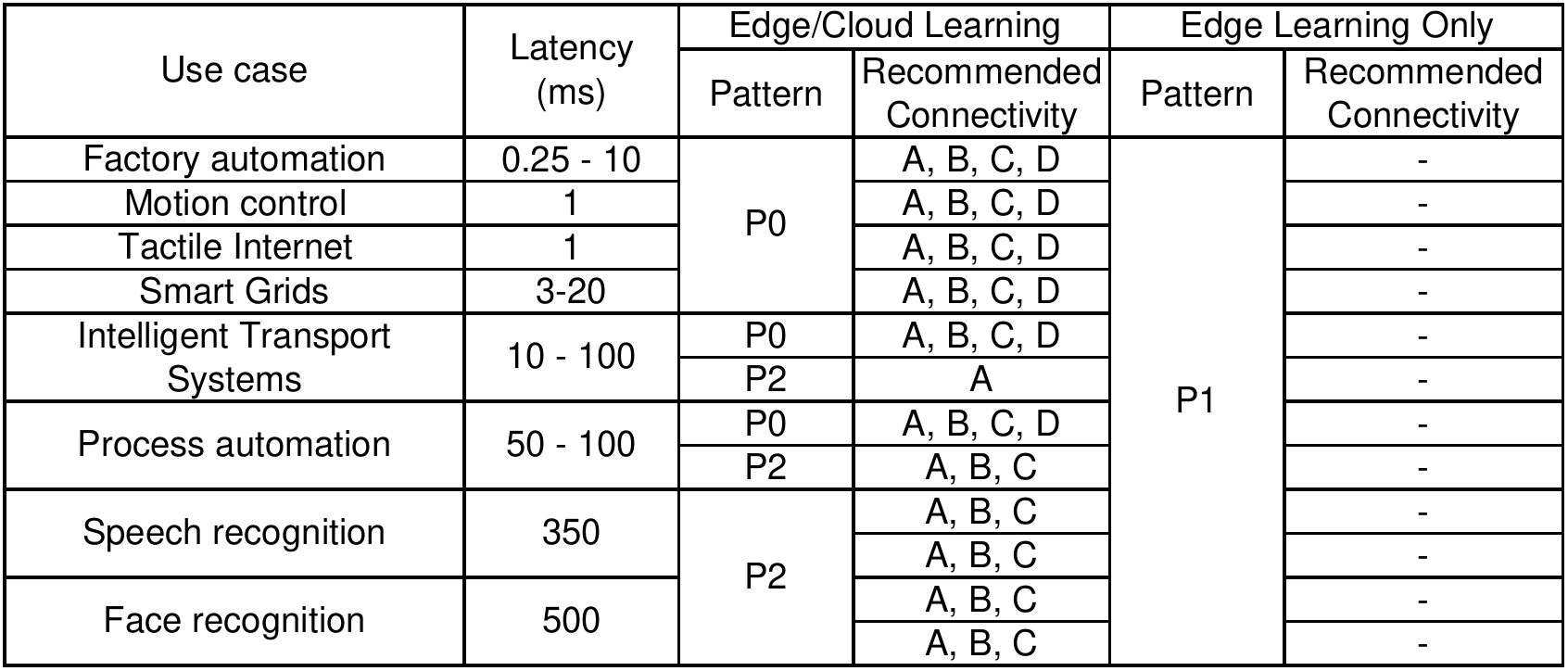}
\end{tabular}
\label{figt:summary}
\end{table}


\section{Summary and future work}
In this paper, we presented the concept of a flow processing stack for IoT that covers the IoT systems on hardware and software layers. We also proposed architectural patterns which, when applied to machine learning algorithms, provide the desired quality i.e. short response times and prediction accuracy.

We showed that the decision to apply delta patterns which involve the distribution of learning process between the edge and the cloud depends on the communication technology, proving that data flows cannot be transformed to executable ensembles without prior analysis of the underlying hardware infrastructure and network topology. Additional results include transfer times and required energy for data sizes of IoT and edge-related tasks, such as sending sensor data and transferring machine learning models. As future work, we plan to evaluate the monitoring metrics for hardware infrastructures and sensor data that would prompt the use of appropriate delta patterns for machine learning. 

\section*{Acknowledgements}
The research presented in this paper was partially supported by the National Centre for Research and Development (NCBiR) under Grant No. LIDER/15/0144/L-7/15/NCBR/2016.

\bibliographystyle{elsarticle-num}
\bibliography{flowstack-bib}

\begin{thebibliography}{10}
\expandafter\ifx\csname url\endcsname\relax
  \def\url#1{\texttt{#1}}\fi
\expandafter\ifx\csname urlprefix\endcsname\relax\def\urlprefix{URL }\fi
\expandafter\ifx\csname href\endcsname\relax
  \def\href#1#2{#2} \def\path#1{#1}\fi

\bibitem{rajesh2010integration}
V.~Rajesh, J.~Gnanasekar, R.~Ponmagal, P.~Anbalagan, Integration of wireless
  sensor network with cloud, in: Recent Trends in Information,
  Telecommunication and Computing (ITC), 2010 International Conference on,
  IEEE, 2010, pp. 321--323.

\bibitem{brzoza2016fpga}
R.~Brzoza-Woch, P.~Nawrocki, {FPGA-Based Web Services--Infinite Potential or a
  Road to Nowhere?}, IEEE Internet Computing 20~(1) (2016) 44--51.

\bibitem{Guinard:2016:BWT:3055920}
D.~Guinard, V.~Trifa, Building the Web of Things: With Examples in Node.Js and
  Raspberry Pi, 1st Edition, Manning Publications Co., Greenwich, CT, USA,
  2016.

\bibitem{morrison2010flow}
J.~P. Morrison, Flow-Based Programming: A new approach to application
  development, CreateSpace, 2010.

\bibitem{bergius2015noflo}
H.~Bergius, Noflo--flow-based programming for javascript, URL: http://noflojs.
  org.

\bibitem{blackstock2012iot}
M.~Blackstock, R.~Lea, Iot mashups with the wotkit, in: Internet of Things
  (IOT), 2012 3rd International Conference on the, IEEE, 2012, pp. 159--166.

\bibitem{Hermann:2016}
M.~Hermann, T.~Pentek, B.~Otto, Design principles for industrie 4.0 scenarios,
  in: System Sciences (HICSS), 2016 49th Hawaii International Conference on,
  IEEE, 2016, pp. 3928--3937.

\bibitem{Brettel:2014}
M.~Brettel, N.~Friederichsen, M.~Keller, M.~Rosenberg, How virtualization,
  decentralization and network building change the manufacturing landscape: An
  industry 4.0 perspective, International Journal of Mechanical, Industrial
  Science and Engineering 8~(1) (2014) 37--44.

\bibitem{4519604}
E.~A. Lee, Cyber physical systems: Design challenges, in: 2008 11th IEEE
  International Symposium on Object and Component-Oriented Real-Time
  Distributed Computing (ISORC), 2008, pp. 363--369.

\bibitem{BALIS2018128}
B.~Balis, R.~Brzoza-Woch, M.~Bubak, M.~Kasztelnik, B.~Kwolek, P.~Nawrocki,
  P.~Nowakowski, T.~Szydlo, K.~Zielinski, Holistic approach to management of it
  infrastructure for environmental monitoring and decision support systems with
  urgent computing capabilities, Future Generation Computer Systems 79~(Part 1)
  (2018) 128 -- 143.

\bibitem{abadi2016tensorflow}
M.~Abadi, P.~Barham, J.~Chen, Z.~Chen, A.~Davis, J.~Dean, M.~Devin,
  S.~Ghemawat, G.~Irving, M.~Isard, et~al., Tensorflow: A system for
  large-scale machine learning., in: OSDI, Vol.~16, 2016, pp. 265--283.

\bibitem{lane2016accelerated}
N.~D. Lane, S.~Bhattacharya, P.~Georgiev, C.~Forlivesi, F.~Kawsar, Accelerated
  deep learning inference for embedded and wearable devices using deepx, in:
  Proceedings of the 14th Annual International Conference on Mobile Systems,
  Applications, and Services Companion, ACM, 2016, pp. 109--109.

\bibitem{bang201714}
S.~Bang, J.~Wang, Z.~Li, C.~Gao, Y.~Kim, Q.~Dong, Y.-P. Chen, L.~Fick, X.~Sun,
  R.~Dreslinski, et~al., 14.7 a 288$\mu$w programmable deep-learning processor
  with 270kb on-chip weight storage using non-uniform memory hierarchy for
  mobile intelligence, in: Solid-State Circuits Conference (ISSCC), 2017 IEEE
  International, IEEE, 2017, pp. 250--251.

\bibitem{BitFusion}
H.~Sharma, J.~Park, N.~Suda, L.~Lai, B.~Chau, V.~Chandra, H.~Esmaeilzadeh,
  \href{https://doi.org/10.1109/ISCA.2018.00069}{Bit fusion: Bit-level
  dynamically composable architecture for accelerating deep neural networks},
  in: Proceedings of the 45th Annual International Symposium on Computer
  Architecture, ISCA '18, IEEE Press, Piscataway, NJ, USA, 2018, pp. 764--775.
\newblock \href {http://dx.doi.org/10.1109/ISCA.2018.00069}
  {\path{doi:10.1109/ISCA.2018.00069}}.
\newline\urlprefix\url{https://doi.org/10.1109/ISCA.2018.00069}

\bibitem{shamili2010malware}
A.~S. Shamili, C.~Bauckhage, T.~Alpcan, Malware detection on mobile devices
  using distributed machine learning, in: Pattern Recognition (ICPR), 2010 20th
  International Conference on, IEEE, 2010, pp. 4348--4351.

\bibitem{gupta2017protonn}
C.~Gupta, A.~S. Suggala, A.~Goyal, H.~V. Simhadri, B.~Paranjape, A.~Kumar,
  S.~Goyal, R.~Udupa, M.~Varma, P.~Jain, Protonn: Compressed and accurate knn
  for resource-scarce devices, in: International Conference on Machine
  Learning, 2017, pp. 1331--1340.

\bibitem{kumar2017resource}
A.~Kumar, S.~Goyal, M.~Varma, Resource-efficient machine learning in 2 kb ram
  for the internet of things, in: International Conference on Machine Learning,
  2017, pp. 1935--1944.

\bibitem{liu2017new}
C.~Liu, Y.~Cao, Y.~Luo, G.~Chen, V.~Vokkarane, Y.~Ma, S.~Chen, P.~Hou, A new
  deep learning-based food recognition system for dietary assessment on an edge
  computing service infrastructure, {IEEE} Trans. Services Computing 11~(2)
  (2018) 249--261.

\bibitem{tran2017mobile}
T.~X. Tran, M.-P. Hosseini, D.~Pompili, Mobile edge computing: Recent efforts
  and five key research directions, MMTC Communications-Frontiers 12~(4) (2017)
  29--34.

\bibitem{IndustryFog4}
G.~{Peralta}, M.~{Iglesias-Urkia}, M.~{Barcelo}, R.~{Gomez}, A.~{Moran},
  J.~{Bilbao}, Fog computing based efficient iot scheme for the industry 4.0,
  in: 2017 IEEE International Workshop of Electronics, Control, Measurement,
  Signals and their Application to Mechatronics (ECMSM), 2017, pp. 1--6.
\newblock \href {http://dx.doi.org/10.1109/ECMSM.2017.7945879}
  {\path{doi:10.1109/ECMSM.2017.7945879}}.

\bibitem{LOSING2017}
V.~Losing, B.~Hammer, H.~Wersing, Incremental on-line learning: {A} review and
  comparison of state of the art algorithms, Neurocomputing 275 (2018)
  1261--1274.

\bibitem{yin2016improved}
S.~Yin, X.~Xie, J.~Lam, K.~C. Cheung, H.~Gao, An improved incremental learning
  approach for kpi prognosis of dynamic fuel cell system, IEEE transactions on
  cybernetics 46~(12) (2016) 3135--3144.

\bibitem{provodin2016fast}
A.~Provodin, L.~Torabi, B.~Flepp, Y.~LeCun, M.~Sergio, L.~D. Jackel, U.~Muller,
  J.~Zbontar, Fast incremental learning for off-road robot navigation, CoRR
  abs/1606.08057.

\bibitem{bonomi2012fog}
F.~Bonomi, R.~Milito, J.~Zhu, S.~Addepalli, Fog computing and its role in the
  internet of things, in: Proceedings of the first edition of the MCC workshop
  on Mobile cloud computing, ACM, 2012, pp. 13--16.

\bibitem{yi2015survey}
S.~Yi, C.~Li, Q.~Li, A survey of fog computing: concepts, applications and
  issues, in: Proceedings of the 2015 Workshop on Mobile Big Data, ACM, 2015,
  pp. 37--42.

\bibitem{schulz17}
P.~Schulz, M.~Matthe, H.~Klessig, M.~Simsek, G.~P. Fettweis, J.~Ansari, S.~A.
  Ashraf, B.~Almeroth, J.~Voigt, I.~Riedel, A.~Puschmann, A.~Mitschele{-}Thiel,
  M.~Muller, T.~Elste, M.~Windisch, Latency critical iot applications in 5g:
  Perspective on the design of radio interface and network architecture, {IEEE}
  Communications Magazine 55~(2) (2017) 70--78.

\bibitem{szydlo:2017}
T.~Szydlo, R.~Brzoza-Woch, J.~Sendorek, M.~Windak, C.~Gniady, Flow-based
  programming for iot leveraging fog computing, in: 2017 IEEE 26th
  International Conference on Enabling Technologies: Infrastructure for
  Collaborative Enterprises (WETICE), 2017, pp. 74--79.

\bibitem{drift:1}
G.~Ditzler, M.~Roveri, C.~Alippi, R.~Polikar, Learning in nonstationary
  environments: A survey, IEEE Computational Intelligence Magazine 10~(4)
  (2015) 12--25.

\bibitem{drift:2}
J.~a. Gama, I.~\v{Z}liobait\.{e}, A.~Bifet, M.~Pechenizkiy, A.~Bouchachia, A
  survey on concept drift adaptation, ACM Comput. Surv. 46~(4) (2014)
  44:1--44:37.

\bibitem{FedCSIS2014252}
T.~Szydlo, P.~Nawrocki, R.~Brzoza-Woch, K.~Zielinski, Power aware mom for
  telemetry-oriented applications using gprs-enabled embedded devices a levee
  monitoring use case, in: M.~P. M.~Ganzha, L.~Maciaszek (Ed.), Proceedings of
  the 2014 Federated Conference on Computer Science and Information Systems,
  Vol.~2 of Annals of Computer Science and Information Systems, IEEE, 2014, pp.
  pages 1059--1064.

\bibitem{codegen}
T.~Szydlo, J.~Sendorek, R.~Brzoza-Woch, Enabling machine learning on resource
  constrained devices by source code generation of the learned models, in:
  Y.~Shi, H.~Fu, Y.~Tian, V.~V. Krzhizhanovskaya, M.~H. Lees, J.~Dongarra,
  P.~M.~A. Sloot (Eds.), Computational Science -- ICCS 2018, Springer
  International Publishing, Cham, 2018, pp. 682--694.

\bibitem{surveyiot}
A.~Al-Fuqaha, M.~Guizani, M.~Mohammadi, M.~Aledhari, M.~Ayyash, Internet of
  things: A survey on enabling technologies, protocols, and applications, IEEE
  Communications Surveys \& Tutorials 17~(4) (2015) 2347--2376.

\end{thebibliography}







\end{document}